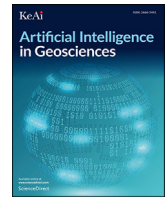

# Local earthquakes detection: A benchmark dataset of 3-component seismograms built on a global scale

Fabrizio Magrini [a,*], Dario Jozinović [a,b], Fabio Cammarano [a], Alberto Michelini [b], Lapo Boschi [c,d,e]

[a] Department of Science, Università Degli Studi Roma Tre, Italy
[b] Istituto Nazionale di Geofisica e Vulcanologia (INGV), Rome, Italy
[c] Dipartimento di Geoscienze, Università Degli Studi di Padova, Italy
[d] Sorbonne Université, CNRS, INSU, Institut des Sciences de La Terre de Paris, ISTeP UMR 7193, F-75005, Paris, France
[e] Istituto Nazionale di Geofisica e Vulcanologia, Bologna, Italy

A R T I C L E   I N F O

A B S T R A C T

Machine learning is becoming increasingly important in scientific and technological progress, due to its ability to create models that describe complex data and generalize well. The wealth of publicly-available seismic data nowadays requires automated, fast, and reliable tools to carry out a multitude of tasks, such as the detection of small, local earthquakes in areas characterized by sparsity of receivers. A similar application of machine learning, however, should be built on a large amount of labeled seismograms, which is neither immediate to obtain nor to compile. In this study we present a large dataset of seismograms recorded along the vertical, north, and east components of 1487 broad-band or very broad-band receivers distributed worldwide; this includes 629,095 3-component seismograms generated by 304,878 local earthquakes and labeled as EQ, and 615,847 ones labeled as noise (AN). Application of machine learning to this dataset shows that a simple Convolutional Neural Network of 67,939 parameters allows discriminating between earthquakes and noise single-station recordings, even if applied in regions not represented in the training set. Achieving an accuracy of 96.7, 95.3, and 93.2% on training, validation, and test set, respectively, we prove that the large variety of geological and tectonic settings covered by our data supports the generalization capabilities of the algorithm, and makes it applicable to real-time detection of local events. We make the database publicly available, intending to provide the seismological and broader sci-entific community with a benchmark for time-series to be used as a testing ground in signal processing.

## 1. Introduction

Natural earthquakes are the shaking of the Earth surface caused by a sudden release of elastic energy from geological faults which generates mechanical waves, called seismic waves. The strength of an earthquake, generally indicated by its magnitude, is proportional to the logarithm of the energy liberated (e.g. Båth, 1955), and determines our ability to perceive the ground motion due to the seismic-wave propagation. Over the last century, the possibility of recording the arrival times of different seismic phases at sensitive instruments (i.e. seismographs) has enabled seismologists to image and understand the Earth's interior and dynamics. Among these seismic phases are the compressional (P) and shear (S) waves, which are generally the first to be recorded at a seismic receiver when an earthquake occurs and should be considered as the two fundamental types of seismic waves, in that they generate all the others (e.g. surface waves) by interacting with the discontinuities within the Earth.

The enhancement and spreading of seismic sensors around the world, together with the theoretical progress made in seismology over the last decades, nowadays allow not only to exploit seismic signals emitted by earthquakes, but also those connected to ambient noise (e.g. Shapiro and Campillo, 2004; Boschi and Weemstra, 2015). This study aims to provide a dataset of labeled seismograms generated by both local earthquakes and noise, and recorded at a large number of seismic receivers distributed around the world (Fig. 4). The choice of collecting only local earthquake-data is motivated by the fact that small-magnitude events, which generate relatively small amplitudes and are easily attenuated, are often problematic to detect but provide valuable information about

---

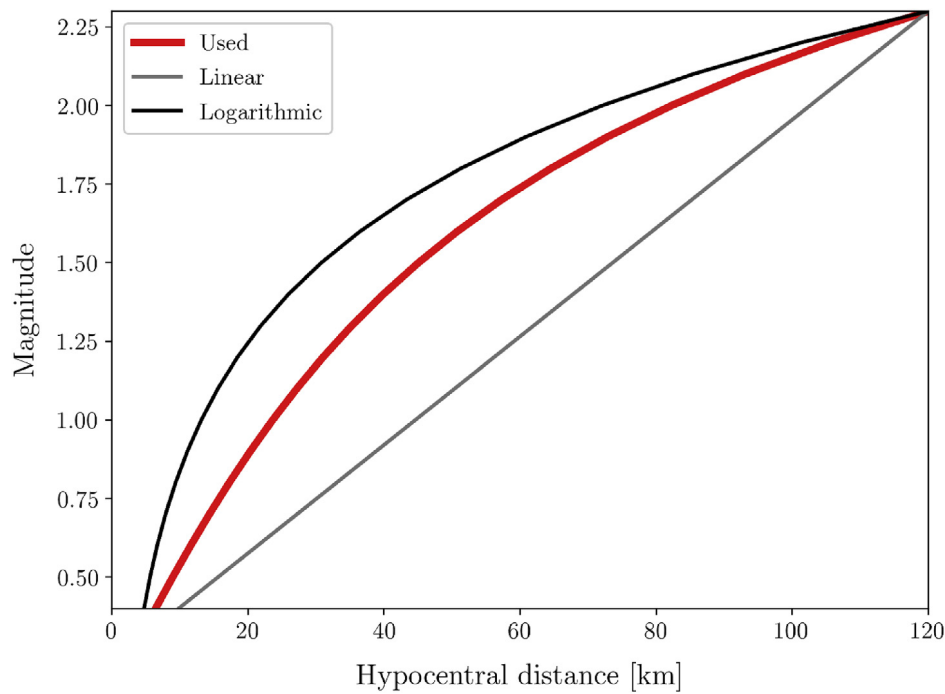

**Fig. 1.** Perceptibility radius used for collecting earthquake-data (red) compared with those obtained by using a logarithmic (black) and a linear (gray) relation between hypocentral distance and magnitude. (For interpretation of the references to color in this figure legend, the reader is referred to the Web version of this article.)

earthquake processes (Brodsky, 2019). Cataloging small earthquakes could be important, for example, for better understanding how earthquakes interact with one another, their reoccurrence, nucleation stage and the foreshock evolution (Ross et al., 2019).

This global dataset is intended to be used for carrying out a multitude of seismological and signal processing tasks on single-station recordings, and its size particularly suits machine learning (ML) applications. ML is becoming increasingly important in scientific and technological progress, due to its ability to create models that describe complex data. In the field of seismology, ML algorithms have proved to be often more reliable than expert scientists in recognizing seismic phases arrivals (e.g. Zhu et al., 2019) and determining physical quantities associated with the earthquake (e.g. Ross et al., 2018). This has important implications, e.g., for the improvement of modern earthquake early-warning system techniques and therefore for the mitigation of risk (Meier et al., 2019). ML applications, however, always require a large number of samples to induce these models to generalize well, i.e. to properly classify data not represented in the training set (for an overview on applications of ML in seismology see, e.g., Kong et al., 2019). At the present time, availability of seismological benchmark datasets like the one presented here is very limited. To our knowledge, the only instance of something similar in size has been assembled and published in a recent, independent study by Mousavi et al. (2019a). The impressive work carried out by these authors, however, led to a dataset of different characteristics, arising from e.g. different processing and geographic distribution of the seismograms collected.

We hope that a collection of time-series like the one presented here may benefit not only seismologists, but a broader community including data scientists interested in informative data such as seismograms recorded on the Earth surface. After explaining the procedure adopted for an automated labeling of the waveforms (Section 2), we describe in detail the features of the dataset (Section 3). An application of supervised ML to a binary classification problem is presented in Section 4. Specifically, we show the ability of a Convolutional Neural Network (Krizhevsky et al., 2012) trained on our dataset to recognize earthquakes from noise in unlabeled data (i.e. test set) based on single-station recordings. Possible applications of the dataset and conclusions are presented in Sections 5 and 6, respectively.

## 2. Labeling and downloads

We searched for seismic data recorded at more than 1500 publicly-available, broad-band or very broad-band seismic stations equipped with sensors oriented along vertical (Z), north (N), and east (E) components. For each receiver, recordings that satisfied some quality criteria explained in the following paragraphs were downloaded, demeaned, detrended, tapered with a 5% cosine-taper, and bandpass-filtered between 0.1 and 5 Hz before deconvolving the instrumental response to physical units (velocity). Each 3-components seismogram was then cut into time-windows of 27 s sampled at 20 Hz, and labeled as earthquake (EQ) or noise (AN) following an automated procedure, presented below.

### 2.1. Earthquakes

To download EQ waveforms we relied on several catalogues of seismic events (see Data & Resources); for each location, we used the catalogue with the largest number of earthquakes reported, and selected only seismic events satisfying 3 quality parameters. Since this study focuses on local earthquakes, we set the (1) *maximum hypocentral distance* of an earthquake with respect to a receiver to 134 km. Events in the catalogue satisfying this condition are subject to a further selection based on a criterion of (2) *time separation* from other events: if the considered event is disturbed by other events that occurred at about the same time in the vicinity of the station, the event is discarded. In practice, we only use those whose origin times are at least 100 s before and 600 s after the closest available events in the catalogue with epicentral distances ≤ 1.7° (~189 km). This conservative choice is motivated by the need of avoiding arrivals of seismic phases from different local events within the same waveforms.

The last requirement for an event on the catalogue to pass the quality selection is determined by a (3) *perceptibility radius* (see Nuttli and Zollweg, 1974, and our implementation detailed below) that is function of both magnitude and hypocentral distance (Fig. 1): for each magnitude, the perceptibility radius indicates the maximum acceptable hypocentral distance above which the event is discarded since no visible signal can likely be detected. The strategy of avoiding such events is motivated by the fact that the capability of a receiver to record an earthquake decreases with the hypocentral distance and strongly depends on the attenuation properties of the Earth, which can be significantly variable depending on e.g. the local geology and/or the tectonic environment (e.g. Dalton et al., 2008; Dalton and Faul, 2010). The choice of the perceptibility radius is critical in compromising the trade-off between the number of rejected events and the quality of the downloaded data; however, accounting for its spatial variability in order to optimize the labeling at each location covered by our dataset would be, at least, impractical. For this reason, our choice of the perceptibility radius is empirical, and has been made after visual inspection of its performance in terms of quality of the labeled waveforms vs. rejection rate. In this regard, we visually checked a large number of seismograms (more than 40,000) and excluded from the dataset those stations which appeared too noisy and therefore did not bring clear evidence of earthquakes in the data. The bad quality of such waveforms can be ascribed either to an inappropriate definition of the perceptibility radius or strong ambient noise at those locations, or to relatively large errors on the events information provided in the catalogues (due to, e.g., a scarce seismic coverage in the surroundings of certain receivers).

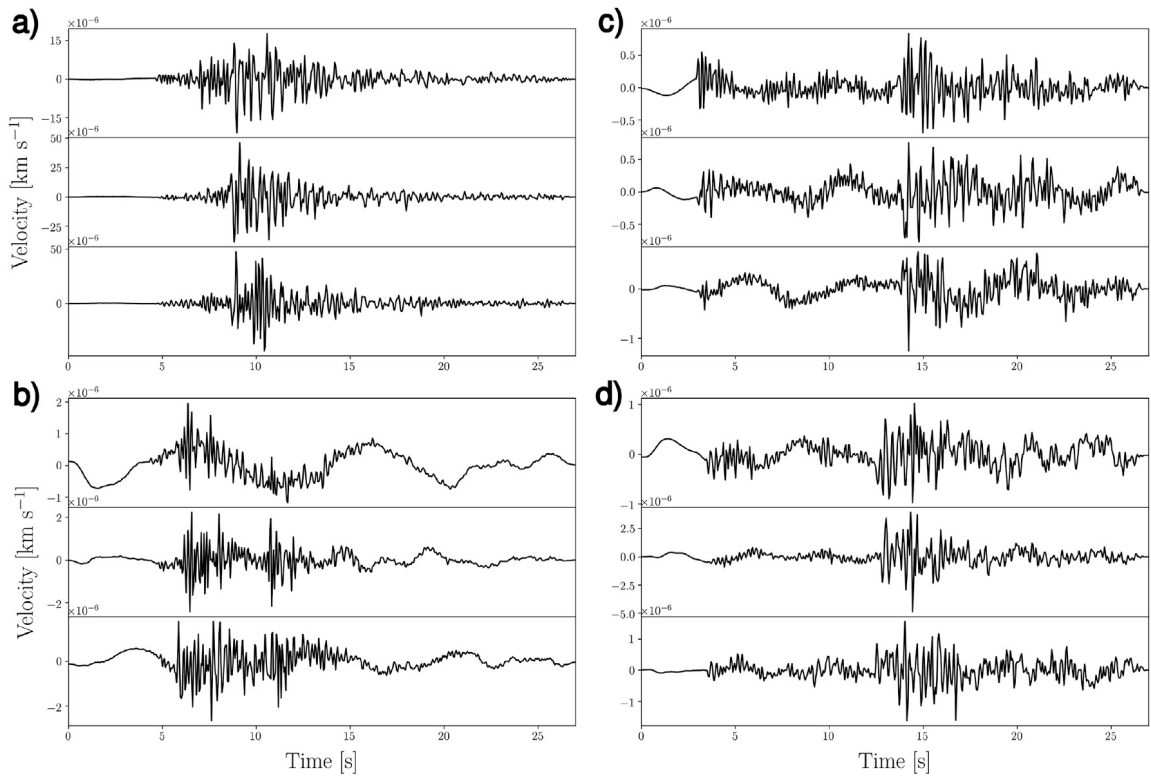

**Fig. 2.** Four randomly selected seismograms (Z, N, E components from top to bottom of each panel) which passed the selection criteria explained in Section 2.1 and were consequently labeled as earthquakes. Each recording brings evidence of a particle motion due to a seismic event. Station codes, start times of the waveforms, origin times and magnitudes of the earthquakes (as indicated by the catalogue providers) are reported in the following. (a) station code: CI.PALA, start time: 2018-10-20 23:16:46, event time: 2018-10-20 23:16:45, magnitude: 2.5; (b) station code: AE.U15A, start time: 2015-03-03 14:18:56, event time: 2015-03-03 14:18:44, magnitude: 2.4; (c) station code: IV.ATPC, start time: 2014-03-03, 10:08:02, event time: 2014-03-03 10:08:04, magnitude: 1.3; (d) station code: CN.MOBC, start time: 2014-06-26 15:15:11, event time: 2014-06-26 15:15:03, magnitude: 2.3.

In practice, the perceptibility radius has been defined using the functions *linspace* and *geomspace* of NumPy Python library (Oliphant, 2006): we chose the maximum acceptable hypocentral distances within an interval between 4 km and 120 km using 21 points (a) spaced evenly, i.e. $Linear = linspace(4, 120, 21)$, and (b) spaced evenly on a logarithmic scale, i.e. $Logarithmic = geomspace(4, 120, 21)$; each of these two arrays has then been associated with a set of 21 magnitudes evenly spaced between 0.3 and 2.3 to obtain the gray line and black curve shown in Fig. 1, respectively. Magnitudes above 2.4 are always accepted provided condition (1) is met. The perceptibility radius employed in this study has been obtained by defining the array of the hypocentral distances as a weighted average of (a) and (b): $Used = \frac{1}{3}(Linear + 2Logarithmic)$ (red curve in Fig. 1).

For each event that met the above conditions, 3-components seismograms starting 4 s before the expected arrival time of the P-wave at the receiver (calculated using IASP91 as 1-D background model, Kennett and Engdahl, 1991) were downloaded and labeled as EQ (Fig. 2).

### 2.2. Noise

Concerning the labeling of noise data (Fig. 3), we followed the same criterion of *time separation* from the closest events reported in the catalogue, already described above: each waveform labeled as AN is randomly downloaded, provided its starting time and ending time are separated from the closest events at least 100 s and 600 s, respectively. It might be noted that this approach would not prevent from labeling as AN recordings of ground motion generated by seismic events at epicentral distances greater than 1.7° and strong enough to be detected. This choice, however, is supported by two considerations. (1) The Gutenber-Richter relation (Gutenberg and Richter, 1944) says that the probability of occurrence of an earthquake decreases, to a good approximation, exponentially with increasing magnitudes; this circumstance alone makes the probability of randomly labeling as AN an earthquake strong enough to be perceptible at the station location relatively small. In addition, (2) the characteristics of a seismogram recording seismic waves generated by a strong, distant earthquake will be substantially different from those of a seismogram produced by a local earthquake. Indeed, the chosen duration for the waveforms (27 s), together with the maximum hypocentral distance (134 km) used for labeling EQ, imply that EQ seismograms will necessarily record both P, S, and surface waves generated by a seismic event, due to simple velocity-time-distance considerations (for average velocities of these seismic phases see, e.g., Stein and Wysession, 2009). However, since these seismic waves travel at different velocities, the same would not happen in case of a more distant earthquake, thus without affecting the possibility of discriminating between noise- and earthquake-waveforms.

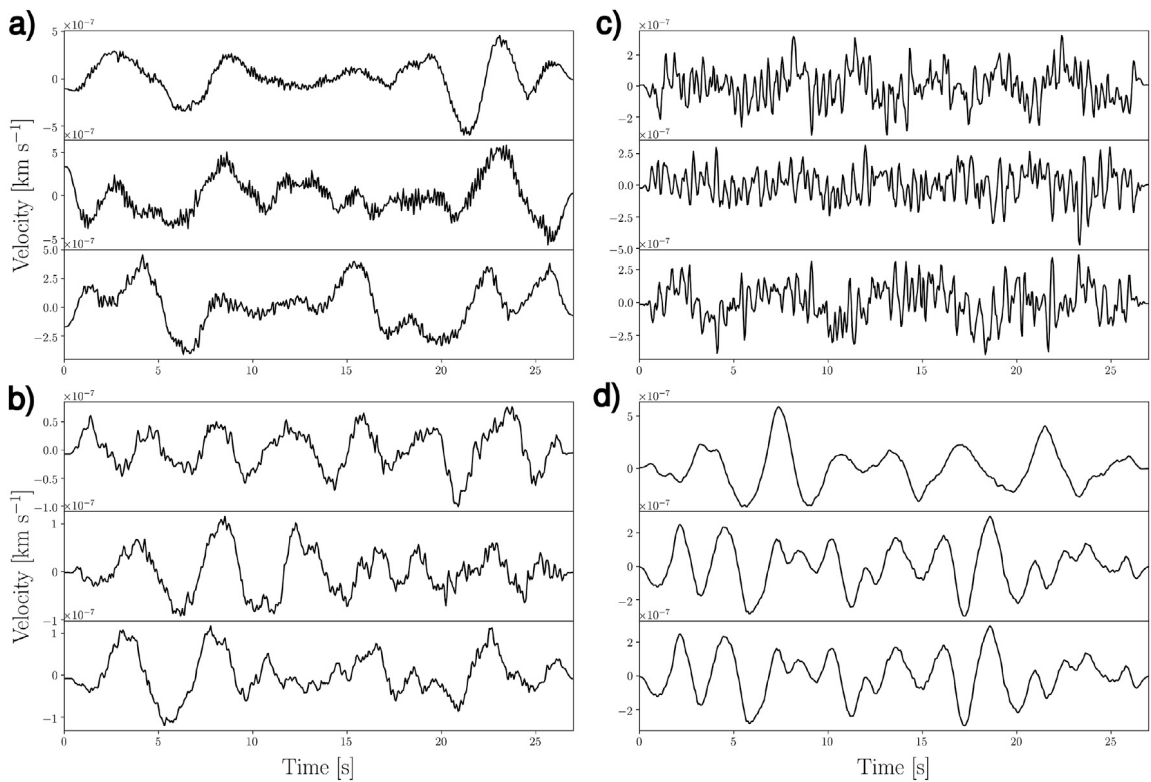

**Fig. 3.** Four randomly selected seismograms (Z, N, E components from top to bottom of each panel) which passed the selection criteria explained in Section 2.2 and were consequently labeled as noise. Station codes and start times of the waveforms are reported in the following. a) station code: OK.ELIS, start time: 2017-05-02 05:46:30; (b) station code: IV.ATMI, start time: 2015-09-11 12:05:38; (c) station code: IV.FIAM, start time: 2018-08-05 20:25:34; (d) station code: FR.TURF, start time: 2016-11-25 22:48:35.

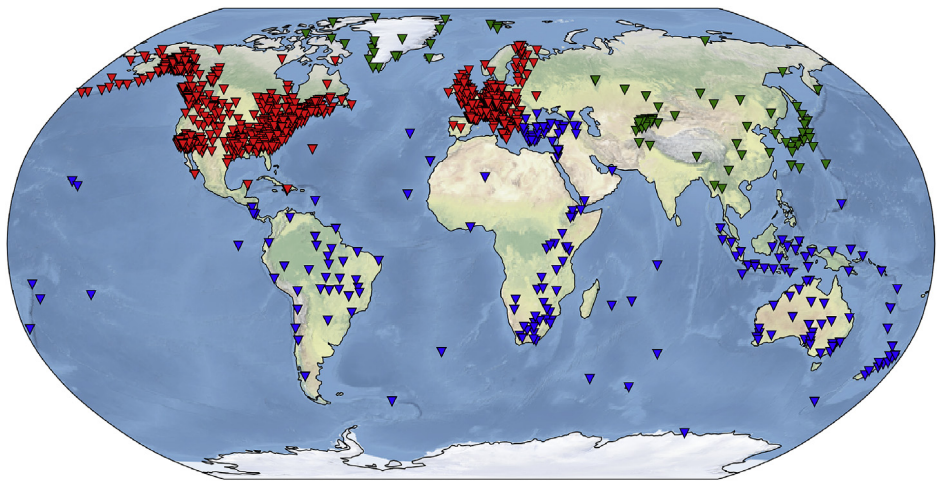

**Fig. 4.** Location of the receivers whose recordings are collected in our dataset. Red, blue, and green indicate training, validation, and test sets as employed in the ML application (Section 4), respectively. (For interpretation of the references to color in this figure legend, the reader is referred to the Web version of this article.)

## 3. Final dataset

The above procedure allowed us to collect 1,244,942 3-component seismograms recorded at 1487 receivers distributed worldwide (Fig. 4): 615,847 labeled as AN and 629,095 as EQ. EQ data have been retrieved from a total amount of 304,878 different earthquakes (Fig. 5), whose magnitude distribution is shown to follow the Gutenberg and Richter (1944) distribution in Fig. 6, at least down to magnitudes of ~2.5. For lower magnitudes, the decrease in the number of earthquakes in our EQ data can be ascribed both to the insufficient completeness of the catalogues and to the conservative choice of the perceptibility radius adopted in the downloads.

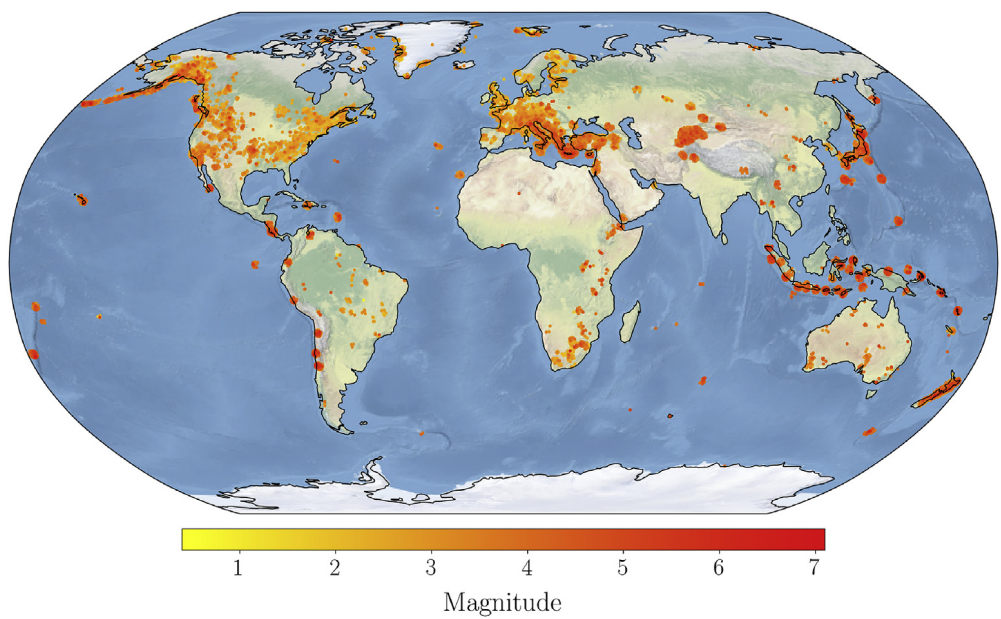

**Fig. 5.** Spatial distribution of the 304,878 earthquakes exploited for collecting 3-component seismograms labeled as EQ.

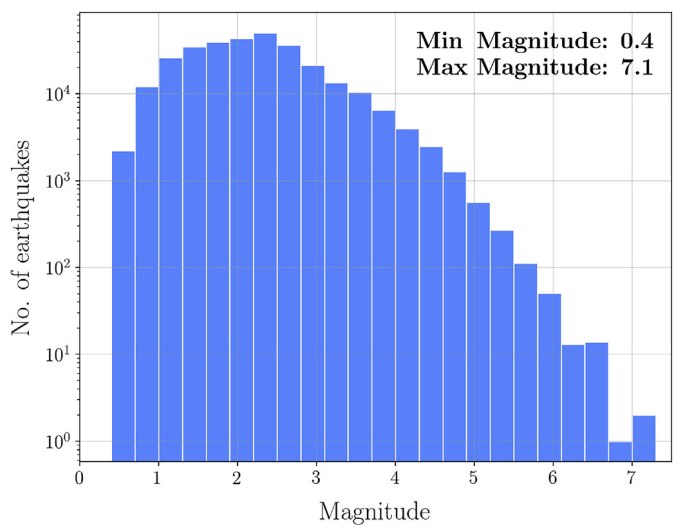

**Fig. 6.** Magnitude distribution of the 304,878 earthquakes exploited for collecting 3-component seismograms labeled as EQ.

We make the dataset publicly available through https://doi.org/10.5281/zenodo.3648232 as a unique file in HDF5 binary data format. Fig. 7 summarizes the structure of the database, which we dubbed LEN-DB (Local Earthquakes and Noise DataBase). The labeled data are split into 2 *HDF5-Groups*: EQ and AN. Each of these groups contains as many *HDF5-Datasets* as the number of 3-component seismograms; these are labeled in accordance to the format *net_sta_starttime*, where *net*, *sta*, and *starttime* represent the seismic network, station, and start time of the seismograms. Each *HDF5-Dataset* (i.e. each triplet of seismograms) has an *attribute*, which allows accessing the respective metadata. Attributes of AN data consist of the station and waveform information: *net* (network code), *sta* (station code), *stla* (station latitude, in degrees, north positive), *stlo* (station longitude, in degrees, east positive), *stel* (station elevation, in meters), *starttime* and *endtime* (start time and end time of the waveforms, respectively); as for EQ data, information about the event are also reported: *mag* (magnitude), *evla* (epicenter latitude, in degrees, north positive), *evlo* (epicenter longitude, in degrees east positive), *evdp* (depth of hypocenter with respect to the nominal sea level given by the WGS84 geoid, in meters), *otime* (event origin time), *dist* (epicentral distance, in meters), *az* (event to station azimuth, in degrees), and *baz* (station to event azimuth, in degrees). In addition, one *HDF5-Group* allows accessing stations' metadata through as many *HDF5-Datasets* as the number of receivers employed for collecting the waveforms.

## 4. Machine learning application

We present in this Section a simple application of the dataset to a signal classification problem. Specifically, we trained a Convolutional Neural Network (CNN) (Krizhevsky et al., 2012) to discriminate between recordings of noise and recordings of earthquakes; the trained model therefore represents a single-station local-earthquake detection algorithm.

### 4.1. Input model

The architecture of the CNN ensures that the algorithm is invariant to a certain degree of data translation and rotation. In other words, when applied to a time-series, the CNN learns its characteristics, regardless of their position in time (Chollet, 2018). The inputs to the CNN are the 27 s 3-component seismograms sampled at 20 Hz. Each input is normalized using the maximum value among the triplet of seismograms, and the maximum is stored and serves as complementary data to the normalized time-series. The architecture of the algorithm is a slightly modified version of ConvNetQuake, a CNN adopted by Perol et al. (2018) and Lomax et al. (2019) for detection and characterization of local and global earthquakes, respectively. The output of the algorithm consists of a real number between 0 and 1, which classifies a given waveform into EQ or AN upon approximation to the closest integer.

The CNN has been set up using the Keras Python library (Chollet et al., 2015). The input layer consists of the normalized waveform array of dimensions (540, 3). This layer is followed by 8 stacked L2-regularized convolutional layers (with regularization constant set to 0.0002), whose number of features is progressively halved by employing max-pooling (e.g. Scherer et al., 2010). The last convolutional layer is flattened and the extracted features, together with the maximum used in the normalization, are then fed to a fully-connected layer with 256 neurons. This configuration resulted in 67,939 model parameters. The rectified linear unit (ReLU) activation function (e.g. Nair and Hinton, 2010) is used throughout the whole architecture except for the last layer, where a fully-connected layer with one neuron returns the classification of the waveform using a sigmoid activation function. Crossentropy (e.g. Goodfellow et al., 2016) and Adam (Kingma and Ba, 2014) are the loss function and the optimization algorithm used throughout the model, respectively.

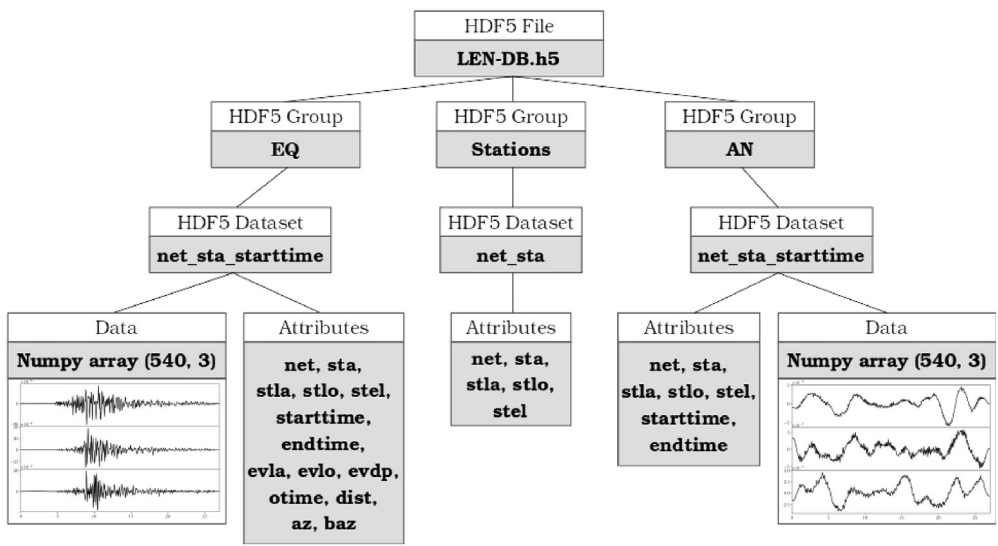

**Fig. 7.** Schematic representation of the structure of the database.

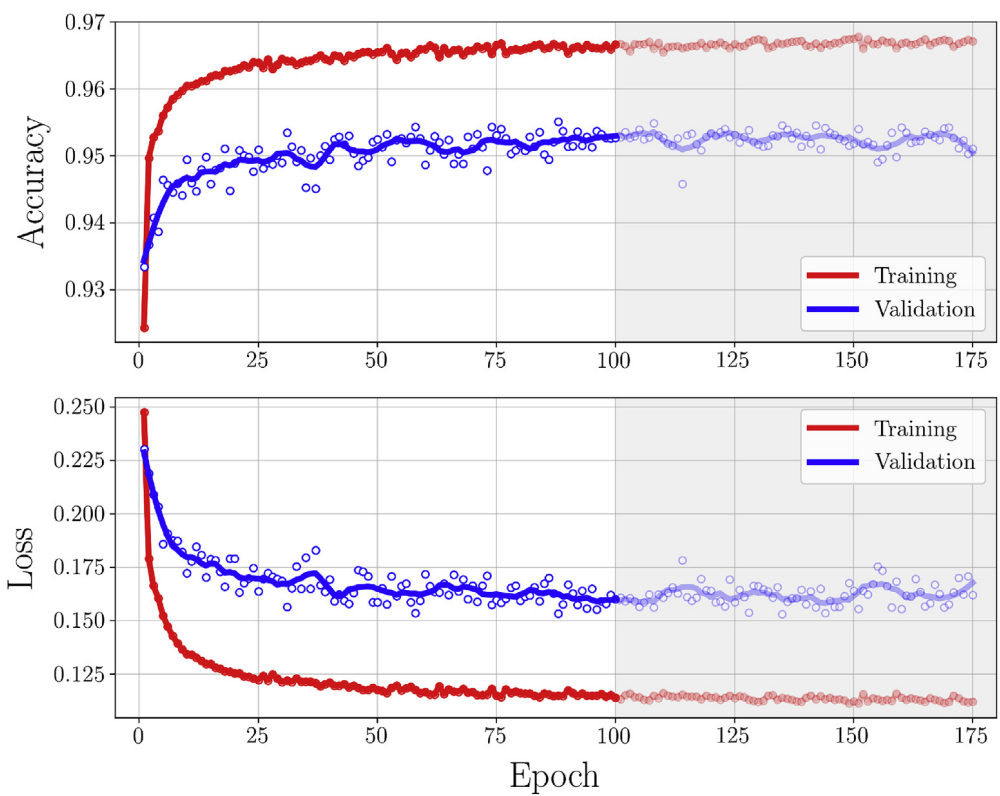

**Fig. 8.** Accuracy (top) and loss (bottom) as function of epoch achieved on training and validation sets. The final model has been trained for 100 epochs, larger epochs are shaded in gray. In each subplot, the red (training) and blue (validation) curves indicate the running average of accuracy/loss actual values (red and blue dots for training and validation, respectively). (For interpretation of the references to color in this figure legend, the reader is referred to the Web version of this article.)

### 4.2. Training and testing

We split the dataset on a geographical basis (Fig. 4), using 884,073 (452,147 EQ, 431,926 AN), 266,407 (128,698 EQ, 137,709 AN), and 94,462 (48,250 EQ, 46,212 AN) 3-components seismograms for training, validation, and test set, respectively; the magnitude distributions of the earthquakes used in these three subsets is illustrated in the supplementary materials. The strategy to split the dataset geographically is adopted to prevent the waveforms of different subsets from carrying geological/tectonic information buried in the signal; in fact, this could possibly induce the model to overfit the data due to information leakage (see, e.g., Chollet, 2018). In other words, the large number of stations from different geological and tectonic settings is probable to make the algorithm learn general features of EQ and AN signals, regardless of the characteristics of the specific areas. Splitting the dataset on a geographical basis therefore helps the generalization capabilities of the model (a thorough discussion on the topic can be found e.g. in Goodfellow et al., 2016). Due to the large number of samples and to limitations in the RAM memory available, for this study we randomly split the training data into three separated subsets. During the training, at each epoch one of them is randomly chosen and used, so that each subset equally contributes to the learning process of the model. The training process required ~30 min on a Nvidia 1060 4 GB for 100 epochs using a batch size of 512 samples. The trained Keras model is available at https://github.com/djozinovi/LEN-DB.

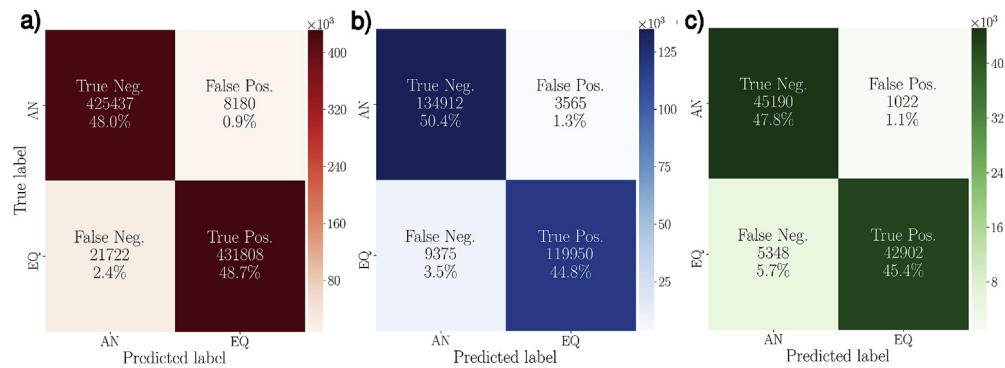

**Fig. 9.** Confusion matrices of (a) train, (b) validation, and (c) test set. In each subpanel, the colorscale indicates the number of 3-component seismograms employed. (For interpretation of the references to colour in this figure legend, the reader is referred to the web version of this article.)

**Table 1**
True Negatives (TN), False Negatives (FN), False Positives (FP), and True Positives (TP) resulting from the application of the detection algorithm to the test set; the results are shown for each seismic network (NET) individually. Accuracy (in percentage) in correctly classifying AN and EQ waveforms are indicated as ACC AN and ACC EQ, respectively, while ACC indicates overall accuracy.

| NET | TN | FN | FP | TP | ACC AN | ACC EQ | ACC |
|-----|------|------|-----|-------|--------|--------|------|
| CN | 184 | 0 | 6 | 66 | 96.8 | 100 | **97.7** |
| DK | 1254 | 32 | 52 | 950 | 96 | 96.7 | **96.3** |
| G | 1291 | 292 | 22 | 1854 | 98.3 | 86.4 | **90.9** |
| GE | 413 | 14 | 13 | 279 | 96.9 | 95.2 | **96.2** |
| IC | 775 | 66 | 48 | 364 | 94.2 | 84.7 | **90.9** |
| II | 4186 | 204 | 65 | 4495 | 98.5 | 95.7 | **97** |
| IU | 2296 | 92 | 45 | 2562 | 98.1 | 96.5 | **97.3** |
| JP | 8438 | 1032 | 77 | 12871 | 99.1 | 92.6 | **95.1** |
| KR | 25227 | 3603 | 684 | 18511 | 97.4 | 83.7 | **91.1** |
| NO | 747 | 9 | 3 | 552 | 99.6 | 98.4 | **99.1** |
| PL | 379 | 4 | 7 | 398 | 98.2 | 99 | **98.6** |

The above procedure yielded an overall accuracy of 96.7% and 95.3% on the training and validation sets, respectively; graphs of accuracy and loss as function of epoch for both training and validation sets are shown in Fig. 8. The performance of the trained model has then been validated over the test set, on which we achieved an overall accuracy of 93.2%.

### 4.3. Discussion of the results

Fig. 9 shows the confusion matrices (e.g. Sammut and Webb, 2017) obtained using our algorithm for classifying the waveforms of the three datasets individually; the percentage of *False Negatives* is larger than the one of *False Positives* for both train, validation, and test sets, albeit small.

This is ascribed to the difficulty of our simple model of 67,939 parameters in detecting earthquakes in presence of relatively high noise levels. To this regard, some seismic networks included in the test set proved to be problematic, contributing to decrease the value of overall accuracy, as illustrated by Table 1. Among them, the KR (Kyrgyzstan) network showed the largest number of undetected earthquakes (i.e. False Negatives). The relatively large number of EQ waveforms belonging to this seismic network turned out to be an important factor in determining the overall accuracy of the subset. In fact, excluding the KR network from the test set allows increasing its overall accuracy from 93.2% to 95.5%, values which are consistent with those achieved on validation and training sets.

Visual inspection of the waveforms misclassified as AN showed that, for the majority of them, the ground motion caused by the earthquake reported on the catalogue is only barely visible, even for magnitudes $\geq$ 3. This is shown in Fig. 10, where only a few of nine randomly selected seismograms recorded by the KR network bring evidence of the earthquake, albeit not as clearly as one would expect from their relatively large magnitudes. This is ascribed to the relatively high noise levels at the locations covered by this seismic network.

The performance of the detection algorithm obtained on individual networks belonging to training and validation sets is shown in Tables 2 and 3, respectively. Except for a few locations, the results confirm the high quality of the labeled seismograms collected in our dataset. It is worth noting the performance of the CNN on the global network IU, which offers an insight into the strong geographic variability of the waveforms; as opposed to a very small number of False Negatives associated with IU at the locations included in test and training sets (see Tables 1 and 2), a relatively poor accuracy is observed for the same network on the validation set (Table 3). In analogy with the KR network, this can be ascribed to high noise levels at specific sensors. On the other hand, the high overall accuracy achieved on the three subsets in presence of such variability of the waveforms with location provides evidence of the good generalization properties of the detection algorithm.

## 5. Possible applications

We have shown in Section 4 that the trained model can be applied to detect small earthquakes in regions that were not represented in the training set. The high accuracy achieved (96.7, 95.3, and 93.2% on training, validation, and test set, respectively) brings evidence that the same method can be applied to real-time detection of earthquakes on individual stations, by streaming continuous data in batches of 27 s on the condition of pre-processing the seismograms as in Section 2. In this regard, a possible attempt to further improve the performance of the algorithm would be introducing one or more recurrent layers in the model (e.g. Mousavi et al., 2019b), to account for the temporal relation between the seismic phases arriving at the receivers.

Our algorithm also suits the analysis of past recordings with the purpose of enriching the actual catalogues by detecting small, local earthquakes that could not be detected by other methods commonly employed (for example the STA/LTA; e.g. Withers et al., 1998); without relying on multiple stations, this would prove especially useful in areas with scarce density of receivers. We have shown that the large variety of geological and tectonic settings covered by our data supports the generalization capabilities of the detection algorithm. On the other hand, for carrying out specific tasks like improving the completeness of the catalogues in certain locations, it might be beneficial to focus only on a portion of the dataset; training a machine learning algorithm on those data would then enable to incorporate the geographic characteristics of the investigated region in the trained model, possibly leading to higher accuracy in the detection. In addition, it might be worth investigating if such a collection of AN data could allow to efficiently simulate seismic ambient noise and extract information on the distribution of noise sources contributing to the recordings of a specific area. Something similar would be particularly useful for constraining the attenuation properties of the region (e.g. Tsai, 2011; Boschi et al., 2019).

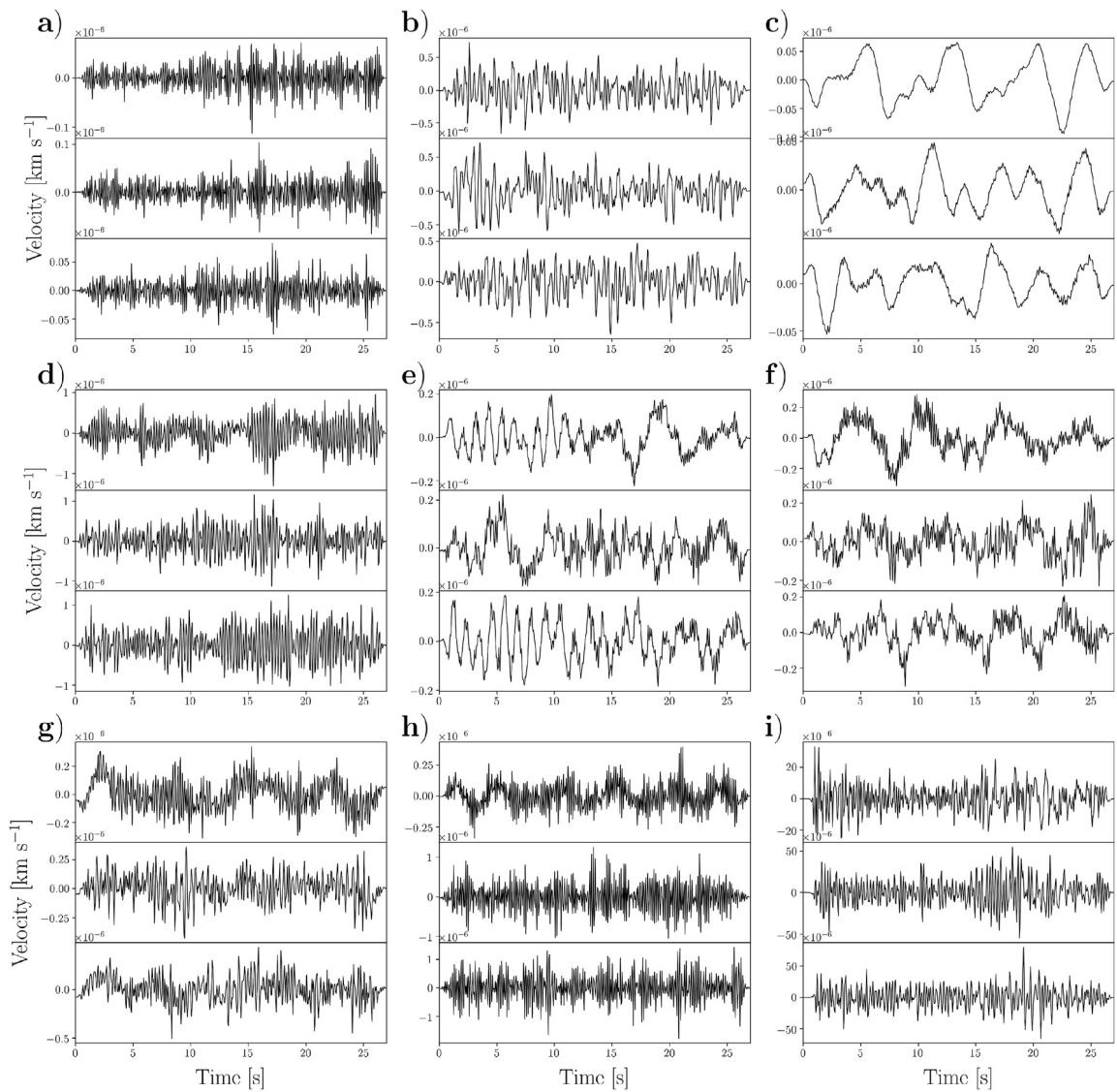

**Fig. 10.** Nine randomly selected seismograms (Z, N, E components from top to bottom of each panel) belonging to the KR network and misclassified as AN. Buried in the seismograms is the evidence of earthquakes characterized by magnitudes ≥ 3. Station codes, start times of the waveforms, origin times and magnitudes of the earthquakes (as indicated by the catalogue providers) are reported in the following. (a) station code: KR.DRK, start time: 2012-04-30 17:43:56, event time: 2012-04-30 17:43:39, magnitude: 3.2; (b) station code: KR.BTK, start time: 2018-08-14 23:05:19, event time: 2018-08-14 23:05:05, magnitude: 3.4; (c) station code: KR.MNAS, start time: 2018-08-25 22:17:44, event time: 2018-08-25 22:17:38, magnitude: 4.9; (d) station code: KR.DRK, start time: 2017-06-20 16:50:17, event time: 2017-06-20 16:50:06, magnitude: 3.0; (e) station code: KR.ANVS, start time: 2018-06-09 23:51:52, event time: 2018-06-09 23:51:42, magnitude: 3.0; (f) station code: KR.ANVS, start time: 2013-09-14 10:00:38, event time: 2013-09-14 10:00:21, magnitude: 3.1; (g) station code: KR.DRK, start time: 2018-08-12 17:18:58, event time: 2018-08-12 17:18:46, magnitude: 3.1; (h) station code: KR.TOKL, start time: 2012-06-10 21:29:00, event time: 2012-06-10 21:28:44, magnitude: 3.2; (i) station code: KR.SFK, start time: 2018-08-08 20:19:47, event time: 2018-08-08 20:19:32, magnitude: 4.3.

Other possible applications of our dataset are connected to signal-processing tasks. Denoising of the waveforms (e.g. Mousavi et al., 2016) and detection of anomalies due to e.g. electronic failures of the sensors are an example; when dealing with real data, it is common to observe the presence of anomalous, meaningless signals which might introduce a bias in the results of a study. In fact, while visually checking the seismograms collected, we noticed the presence of a few instances of such anomalies; although we estimated the amount of such signals to be very small in comparison to the number of healthy waveforms in our dataset (< 0.1%), we tried to remove them employing a robust criterion based on a clustering analysis. However, this procedure (explained in detail in the supplementary materials) did not yield stable results and we decided to keep those anomalies within the dataset, leaving such a task to be tackled by future studies and different methodological approaches.

**Table 2**
Same as Table 1, but obtained on the training set.

| NET | TN | FN | FP | TP | ACC AN | ACC EQ | ACC |
|-----|-----|-----|-----|-----|-----|-----|-----|
| AE | 642 | 6 | 7 | 282 | 98.9 | 97.9 | **98.6** |
| AK | 21315 | 921 | 760 | 28596 | 96.6 | 96.9 | **96.7** |
| CH | 39171 | 2212 | 431 | 34417 | 98.9 | 94 | **96.5** |
| CI | 26044 | 2506 | 581 | 46416 | 97.8 | 94.9 | **95.9** |
| CN | 25973 | 625 | 466 | 22654 | 98.2 | 97.3 | **97.8** |
| CZ | 3666 | 161 | 42 | 3957 | 98.9 | 96.1 | **97.4** |
| FN | 1468 | 35 | 13 | 1150 | 99.1 | 97 | **98.2** |
| FR | 55384 | 5902 | 585 | 48631 | 99 | 89.2 | **94.1** |
| G | 520 | 42 | 1 | 425 | 99.8 | 91 | **95.6** |
| GB | 1865 | 13 | 8 | 596 | 99.6 | 97.9 | **99.2** |
| GE | 2807 | 116 | 58 | 1582 | 98 | 93.2 | **96.2** |
| GR | 5114 | 235 | 67 | 3758 | 98.7 | 94.1 | **96.7** |
| HE | 3028 | 39 | 14 | 2910 | 99.5 | 98.7 | **99.1** |
| II | 8257 | 151 | 41 | 11074 | 99.5 | 98.7 | **99** |
| IU | 6795 | 57 | 98 | 7308 | 98.6 | 99.2 | **98.9** |
| IV | 116349 | 5402 | 3120 | 119243 | 97.4 | 95.7 | **96.5** |
| MN | 22572 | 1497 | 300 | 20280 | 98.7 | 93.1 | **96** |
| N4 | 10555 | 127 | 255 | 7698 | 97.6 | 98.4 | **98** |
| NO | 1489 | 22 | 2 | 901 | 99.9 | 97.6 | **99** |
| OE | 12630 | 234 | 148 | 12813 | 98.8 | 98.2 | **98.5** |
| OK | 41179 | 1084 | 731 | 38321 | 98.3 | 97.2 | **97.8** |
| PE | 954 | 35 | 15 | 450 | 98.5 | 92.8 | **96.6** |
| PL | 620 | 49 | 0 | 676 | 100 | 93.2 | **96.4** |
| UM | 2902 | 11 | 46 | 3644 | 98.4 | 99.7 | **99.1** |
| US | 10264 | 153 | 253 | 8330 | 97.6 | 98.2 | **97.9** |
| XV | 3874 | 87 | 138 | 5696 | 96.6 | 98.5 | **97.7** |

**Table 3**
Same as Tables 1 and 2, but obtained on the validation set.

| NET | TN | FN | FP | TP | ACC AN | ACC EQ | ACC |
|-----|-----|-----|-----|-----|-----|-----|-----|
| AF | 1807 | 75 | 41 | 1369 | 97.8 | 94.8 | **96.5** |
| AU | 4027 | 290 | 107 | 1444 | 97.4 | 83.3 | **93.2** |
| BR | 598 | 4 | 2 | 215 | 99.7 | 98.2 | **99.3** |
| G | 5791 | 733 | 207 | 5327 | 96.5 | 87.9 | **92.2** |
| GE | 52076 | 1771 | 1674 | 44331 | 96.9 | 96.2 | **96.5** |
| GT | 264 | 12 | 6 | 63 | 97.8 | 84 | **94.8** |
| II | 1782 | 47 | 28 | 1509 | 98.5 | 97 | **97.8** |
| IU | 27099 | 4374 | 753 | 21619 | 97.3 | 83.2 | **90.5** |
| MN | 6208 | 44 | 137 | 6689 | 97.8 | 99.3 | **98.6** |
| NZ | 22626 | 1640 | 407 | 25024 | 98.2 | 93.8 | **95.9** |
| TU | 12634 | 385 | 203 | 12360 | 98.4 | 97 | **97.7** |

## 6. Conclusions

We compiled a large dataset of seismograms recorded along the vertical, north, and east components of 1487 broad-band or very broadband receivers distributed worldwide, including 629,095 3-component seismograms generated by 304,878 local earthquakes and labeled as EQ, and 615,847 ones labeled as noise (AN). We used the dataset to train a Convolutional Neural Network (CNN) in discriminating noise-from earthquake-data, and showed that the trained model can be applied to detect small earthquakes in regions that were not represented in the training set. The high accuracy achieved (96.7, 95.3, and 93.2% on training, validation, and test set, respectively) confirm both the high quality of the labeled seismograms collected in the dataset and the good generalization properties of the detection algorithm.

Availability of similar benchmark datasets is, at the present time, very limited. To our knowledge, the only instance of something comparable in size and built on a global scale has been published in a recent, independent work by Mousavi et al. (2019a); differences, however, arise from the distribution of the receivers, processing, and duration of the 3-component seismograms. Our global dataset is intended to be used for carrying out a multitude of seismological and signal processing tasks based on single-station recordings; importantly, its size suits machine learning applications. For this reason, we believe that our large collection of waveforms will not only benefit seismologists, but a broader community including data scientists interested in informative data such as seismograms recorded on the Earth surface.

### 6.1. Data and resources

Catalogues of seismic events were downloaded from Istituto Nazionale di Geofisica e Vulcanologia (INGV) Seismological Data Centre (2006), International Seismological Centre (2019) (Storchak et al., 2013, 2015; Giacomo et al., 2018), and IRIS Data Services (0000). The facilities of IRIS Data Services, and specifically the IRIS Data Management Center, were used for access to waveforms, related metadata, and/or derived products used in this study. IRIS Data Services are funded through the Seismological Facilities for the Advancement of Geoscience and Earth-Scope (SAGE) Proposal of the National Science Foundation under Cooperative Agreement EAR-1261681. Seismic waveforms have been downloaded using EIDA archive (http://www.orfeus-eu.org/eida) from the following network operators: Arizona Geological Survey (2007), Penn State University: AfricaArray (2004), Alaska Earthquake Center (1987), Swiss Seismological Service (SED) at ETH Zurich (1983), California Institute of Technology and United States Geological Survey Pasadena (1926), Geological Survey of Canada (1980), Institute of Geophysics, Academy of Sciences of the Czech Republic (1973), RESIF - Réseau Sismologique et géodésique Français (1995), Institut de Physique Du Globe de Paris (IPGP) & Ecole Et Observatoire Des Sciences De La Terre De Strasbourg (EOST) (1982), GEOFON Data Centre (1993), Federal Institute for Geosciences and Natural Resources (BGR) (1976), Albuquerque Seismological Laboratory (ASL)/USGS (1988, 1990, 1992, 1993), The Finnish National Seismic Network. GFZ Data Services (1980), Scripps Institution of Oceanography (1986), Istituto Nazionale di Geofisica e Vulcanologia (INGV) Seismological Data Centre (2006), Kyrgyz Institute of Seismology, KIS (2007), MedNet Project Partner Institutions (1990), UC San Diego: Central and Eastern US Network (2013), ZAMG - Zentralanstalt für Meterologie und Geodynamik (1987), Oklahoma Geological Survey: Oklahoma Seismic Network (1978), Penn State University: Pennsylvania State Seismic Network (2004), University Of Montana: University of Montana Seismic Network (2017), International Federation of Digital Seismograph Networks: XV Seismic Network (2014) (Tape and West, 2014; Tape et al., 2018).

## Acknowledgments

We are grateful to three anonymous reviewers for their insightful and constructive reviews. We thank the makers of Obspy (Beyreuther et al., 2010). Graphics were created with Python Matplotlib (Hunter, 2007). The Grant to Department of Science, Roma Tre University (MIUR-Italy Dipartimenti di Eccellanza, ARTICOLO 1, COMMI 314 - 337 LEGGE 232/2016) is gratefully acknowledged.

## Appendix A. Supplementary data

Supplementary data to this article can be found online at https://doi.org/10.1016/j.aiig.2020.04.001.